\documentclass[12pt]{article}
\usepackage[english,german,french,polish]{babel}
\usepackage[T1]{fontenc}
\usepackage{amsfonts}

\begin{document}

\selectlanguage{english}

\baselineskip 0.73cm
\topmargin -0.4in
\oddsidemargin -0.1in

\let\ni=\noindent

\renewcommand{\thefootnote}{\fnsymbol{footnote}}

\newcommand{\SM}{Standard Model }

\newcommand{\SMo}{Standard-Model }

\pagestyle {plain}

\setcounter{page}{1}

%\pagestyle{empty}

%\addtocounter{equation}{+1}

~~~~~~
\pagestyle{empty}

\begin{flushright}
IFT-- 11/2
\end{flushright}

\vspace{1.0cm}

{\large\centerline{\bf Hypothesis of photonic portal in the problem}}

{\large\centerline{\bf of dark matter and classical nature of gravitation}}

\vspace{0.5cm}

{\centerline {\sc Wojciech Kr\'{o}likowski}}

\vspace{0.3cm}

{\centerline {\it Institute of Theoretical Physics,  Faculty of Physics, University of Warsaw}}

{\centerline {\it Ho\.{z}a 69, 00--681 Warszawa, ~Poland}}

\vspace{0.6cm}

{\centerline{\bf Abstract}}

\vspace{0.2cm}

Within our model of "photonic portal"\, to the hypothetical hidden sector  of the Universe, an option  is 
considered, where the massive sterile bosons, mediating in this sector and described by the antisymmetric-tensor field $A_{\mu \nu}$ (of dimension one), are gauged by a vector field $\chi_{\mu}$ (of dimension zero). When the 
Einsteinian gravity is switched on into this model, then it is argued that the metric $g_{\mu \nu}(x)$ of spacetime
is proportional to $<\!\!\chi_\mu(x) \chi_\nu(x)\!\!>_{\rm vac}$, being in this way classical in its nature, not the 
subject of quantization within an overall quantum theory including gravitation. A Lagrangian density  (generally 
dependent on $\chi_\mu$ and its first and second derivatives) is tentatively proposed in such a form, where 
$\chi_\mu$ appears also outside of $A_{\mu \nu}$, spoiling the trivial gauge invariance of our earlier model with 
respect to the gauging field $\chi_\mu$. If the cosmological constant $\lambda$ is neglected, then the term with 
$\chi_\mu$ outside of $A_{\mu \nu}$ vanishes. 
 
\vspace{0.4cm}

\ni PACS numbers: 14.80.-j , 04.50.+h , 95.35.+d 

\vspace{0.4cm}

\ni May  2011

\vfill\eject
 
\pagestyle {plain}

\setcounter{page}{1}

\vspace{0.3cm} 

Recently, we speculated [1] on the existence of a hidden sector of the Universe, consisting of some sterile particles (\SMo neutral), {\it e.g.} of  spin-0 bosons ("\,$\!$sterons"), spin-1/2 fermions ("\,$\!$sterinos")  and  nongauge mediating bosons ("$A$ bosons") described by an antisymmetric-tensor field $A_{\mu \nu}$ (of dimension one).   The $A$ bosons were assumed to be weakly coupled to steron-photon pairs and antisterino-sterino pairs by the interaction Lagrangian densities

%rownanie 1
\begin{equation}
- \frac{1}{2} \sqrt{\!f\,}\varphi F_{\mu \nu} A^{\mu \nu}
\end{equation}

\ni and

%rownanie 2
\begin{equation}
- \frac{1}{2} \zeta \sqrt{\!f\,} \bar\psi \,\sigma_{\mu \nu} \psi A^{\mu \nu}\,,
\end{equation}

\ni respectively. Here, $F_{\mu \nu} = \partial_\mu A_\nu - \partial_\nu A_\mu $ is the \SMo electromagnetic field (of dimension two), while $\sqrt{\!f\,}$ and $\zeta \sqrt{\!f\,}$ denote two dimensionless small coupling constants. We  assumed that  

%rownanie 2
\begin{equation}
\varphi = <\!\!\varphi\!\!>_{\rm vac}\! + \,\varphi_{\rm ph}\,,
\end{equation}
 
\ni where $<\!\!\varphi\!\!>_{\rm vac}\, \neq 0$ is a spontaneously nonzero vacuum expectation value of the steron field $\varphi$. The kinetic Lagrangian density of $A$ bosons is built up only from $A_{\mu \nu}$ (of dimension one):

%rownanie 4
\begin{equation}
-\frac{1}{4}\left[\left(\partial_\rho A_{\mu \nu}\right) \left(\partial^\rho  A^{\mu \nu}\right) - M^2 A_{\mu \nu} A^{\mu \nu}\right]\,
\end{equation}

\ni with $M$ standing for a mass scale of $A$ bosons, expected to be large. 

The field equations for $F_{\mu \nu}$ and $A_{\mu \nu}$ , provided by the interactions (1) and (2), are

%rownanie 5
\begin{equation}
\partial^\nu\! \left[F_{\mu \nu} +  \sqrt{\!f\,}\varphi A_{\mu \nu}\right] = -j_\mu \;\;,\;\; F_{\mu \nu} = \partial_\mu A_\nu - \partial_\nu A_\mu 
\end{equation}

\vspace{-0.1cm}

\ni and

\vspace{-0.3cm} 

%rownanie 6
\begin{equation}
(\Box - M^2)A_{\mu \nu} = - \sqrt{\!f\,}\left(\varphi F_{\mu \nu} + \zeta\, \bar\psi\, \sigma_{\mu \nu}\, \psi\right)\,, 
\end{equation}

\ni  where $j_\mu$ is the \SMo electric current. Eqs. (5) give us the form assumed by Maxwell's equations in the presence of our hidden-sector. 

In the first Eq. (5) the total electric source-current is

\vspace{-0.3cm}

%rownanie 7
\begin{equation}
j_\mu + \partial^\nu( \sqrt{\!f\,}\,\varphi A_{\mu \nu})\,,
\end{equation}

\vspace{-0.2cm}

\ni where the second term does not contribute to the total electric charge because of its four-divergence structure:

\vspace{-0.3cm}

%rownanie 8 
\begin{equation}
\int d^3x\,[j_0 + \partial^k(\sqrt{\!f\,}\varphi A_{0 k})] = \int d^3x j_0 = Q\,,
\end{equation}

\vspace{-0.1cm}

\ni so it gets a magnetic character.

We called the coupling (1) of photons to the hidden sector of the Universe "photonic portal"\, (to the hidden sector), in contrast to the popular "Higgs portal"\, [2]. Such a hidden sector is conjectured to be responsible for the cold dark matter [1].

If there exists a vector field $\chi_\mu$ (of dimension zero) gauging the sterile mediating field $A_{\mu \nu}$ (of dimension one),

\vspace{-0.2cm}

%rownanie 9
\begin{equation}
A_{\mu \nu} = \partial_\mu \chi_\nu - \partial_\nu \chi_\mu \,,
\end{equation}

\vspace{-0.1cm}

\ni ({\it cf.} Appendix in Ref. [3]) and if, in addition, $\chi_\mu$ still does not appear outside of $A_{\mu \nu}$ in the total Lagrangian (nor in a subsidiary condition), our model of hidden sector remains essentially unchanged {\it versus} the previous nongauge case (and is then trivially gauge invariant with respect to $\chi_\mu$). It is the matter of experiment to decide, whether  the field $\chi_\mu$ (when it exists) is or is not trivial in the above sense. 

When the Einsteinian gravity is switched on in our model, terms including $\chi_\mu$  may appear also outside of

\vspace{-0.2cm}

%rownanie 10
\begin{equation}
A_{\mu \nu} =  \chi_{\nu ; \mu}  - \chi_{\mu ; \nu} \equiv \partial_\mu \chi_\nu -\partial_\nu \chi_\mu
\end{equation}

\vspace{-0.1cm}

\ni in the total Lagrangian and so in the field equation for $\chi_\mu$ as well. Assume tentatively the following Lagrangian density for the field $\chi_\mu$:
 
\vspace{-0.2cm}

%rownanie 11
\begin{eqnarray}
\cal{L}^{(\chi)} & = &- \frac{1}{4}\left(A_{\mu \nu ;\rho} A^{\mu \nu ;\rho} - M^2 A_{\mu \nu}A^{\mu \nu} \right) - \frac{1}{2}\sqrt{f\,}\left(\varphi F_{\mu \nu} + \zeta\, \bar\psi\, \sigma_{\mu \nu}\, \psi\right) A^{\mu \nu} \nonumber \\ & - & \frac{1}{2} \xi M^2\left[\left( R_{\mu \nu} - \frac{1}{2} g_{\mu \nu} R^\rho_\rho\right)  + 8\pi G_N T_{\mu \nu} \right] \chi^\mu \chi^\nu \,. 
\end{eqnarray}

%\vspace{-0.1cm}

\ni Here, $\!(...)_{;\rho}$ denotes covariant differentiation with respect to $x^\rho\!$ and $\!(...)^{;\rho} \!\equiv\! g^{\rho \sigma}(...)_{;\sigma}$ , $G_N \!=\! 1/M^2_P$ ($\hbar = 1 = c$) stands for the Newton gravitational constant and $\xi$ represents a new dimensionless parameter, while $R_{\mu \nu}  \equiv R^\rho_{\;\mu \nu \rho}$ is the Ricci curvature tensor of spacetime and $T_{\mu \nu}$ -- the total energy-momentum tensor of all fields of "matter".

If Einstein gravitational equation

%rownanie 12
\begin{equation}
R_{\mu \nu} - \frac{1}{2} g_{\mu \nu} R^\rho_\rho -\lambda g_{\mu \nu} = - 8\pi G_N T_{\mu \nu}   
\end{equation}

\ni or equivalently 

%rownanie 13
\begin{equation}
R_{\mu \nu}+ \lambda g_{\mu \nu} = - 8\pi G_N\left( T_{\mu \nu}  - \frac{1}{2} g_{\mu \nu} T^\rho_\rho\right)   
\end{equation}

\ni  is taken into account, then the Lagrangian density (11) reads

%rownanie 14
\begin{equation}
{\cal{L}}^{(\chi)} = - \frac{1}{4}\left(A_{\mu \nu ;\rho} A^{\mu \nu ;\rho} - M^2 A_{\mu \nu}A^{\mu \nu} \right) - \frac{1}{2}\sqrt{f\,}\left(\varphi F_{\mu \nu} + \zeta\, \bar\psi\, \sigma_{\mu \nu}\, \psi\right) A^{\mu \nu} - 
\frac{1}{2}\xi\lambda M^2 \chi_\mu \chi^\mu\,. 
\end{equation}

\ni Here, $\lambda$ is the cosmological constant (of dimension two). If $\lambda \neq 0$, the Lagrangian density (14) is not gauge invariant with respect to the gauging field $\chi_\mu$, as it is trivially gauge invariant without gravitation. Without gravitation, the Lagrangian density (14) is reduced to the form 

%rownanie 15
\begin{equation}
{\cal{L}}^{(\chi)} = {\cal{L}}^{(A)} = - \frac{1}{4}\left[\left(\partial_\rho A_{\mu \nu}\right) \left( \partial^\rho A^{\mu \nu}\right) - M^2 A_{\mu \nu}A^{\mu \nu} \right] - \sqrt{f\,}\left(\varphi F_{\mu \nu} + \zeta\, \bar\psi\, \sigma_{\mu \nu}\, \psi\right) A^{\mu \nu} \,
\end{equation}

\ni equal to the sum of Eqs. (1), (2) and (4).

In the Einstein equation (12) or (13), the total energy-momentum tensor of "matter"$\,$ is given as $(1/2)\sqrt{-g\,} T_{\mu \nu} = \partial\left(\sqrt{-g\,} \cal{L} \right)/\partial g^{\mu \nu} - \partial_\rho {\partial \left(\sqrt{-g\,} \cal{L}\right)/ \partial} \partial_\rho g^{\mu \nu} $, where $\cal{L}$ denotes the total Lagrangian density of "matter"\,($\cal{L} = \cal{L}^{(\chi)} + ...$) and $g = {\rm det}(g_{\mu \nu})$. Then, the variational principle $\delta_g \int d^4 V \sqrt{-g\,}[(R+2\lambda)/(16\pi G_N) + {\cal{L}}] = 0$ with $R = R^\rho_\rho $ leads to Eq. (12) or (13).

Note that in Eqs. (11) and (12) or (13) [4]

%rownanie 16
\begin{equation}
R_{\mu\nu}  \equiv R^\rho\,_{\mu\nu\rho} \equiv \Gamma^\rho_{\mu \rho,\nu} - \Gamma^\rho_{\mu\nu,\rho} + \Gamma^\sigma_{\mu\rho} \Gamma^\rho_{\nu\sigma} - \Gamma^\sigma_{\mu\nu}\Gamma^\rho_{\rho\sigma}
\end{equation}

\ni with $(...)_{,\nu} \equiv \partial_\nu(...)$, giving the identity

%rownanie 17
\begin{equation}
\chi^\nu\!\,_{;\mu ;\nu} - \chi^\nu\!\,_{;\nu ;\mu} \equiv - R_{\mu\nu} \chi^\nu\,. 
\end{equation}

\ni Here,

%rownanie 18
\begin{equation}
\Gamma^\rho_{\mu \nu} \equiv  \frac{1}{2} g^{\rho \sigma}\left(g_{\sigma\mu,\nu} + g_{\sigma\nu,\mu} - g_{\mu\nu,\sigma}\right) 
\end{equation}

\ni and {\it e.g.}

%rownanie 19
\begin{equation}
\chi_{\mu ;\nu}  \equiv  \chi_{\mu,\nu} - \Gamma^\rho_{\mu\nu}\chi_\rho \;\,,\;\,\chi^{\mu}_{\;\, ;\nu}  \equiv  \chi^{\mu}_{\;\,,\nu} + \Gamma^\mu_{\nu \rho}\chi^\rho \,.
\end{equation}

\ni In Eq. (12), $g_{\mu\nu}^{\;\;\;\, ;\nu} =0$ because of $g_{\mu\nu}^{\;\;\;\, ;\rho} =0$ and $(R_{\mu\nu} -\frac{1}{2}g_{\mu\nu}R^\rho_\rho)^{;\nu} = 0$ due to the Bianchi identities, thus $T_{\mu\nu}^{\;\;\, ;\nu} =0$.

The field equation for $\chi_\mu$ following as an Euler-Lagrange equation from the variational principle $\delta_\chi\int d^4V \sqrt{-g\,}{\cal{L}}^{(\chi)} = 0$ with the tentative Lagrangian density (14) (dependent on $\chi_\mu\,,\,A_{\mu \nu}$ and $A_{\mu \nu ;\rho}$) is 

%rownanie 20
\begin{equation}
-(A_{\mu \nu ;\rho}^{\;\;\;\;\;\;;\rho} \!+\!\frac{1}{\sqrt{-g\,}} g^{\sigma \lambda}\Gamma^\rho_{\sigma \lambda}A_{\mu \nu;\rho} \!+\! M^2 A_{\mu \nu})^{;\nu} \!\!+ \xi \lambda M^2\chi_\mu = -\sqrt{f\,}(\varphi F_{\mu \nu} \!+\! \zeta\, \bar\psi\, \sigma_{\mu \nu} \psi)^{;\nu}.
\end{equation}

\ni Here, $g^{\sigma \lambda}\Gamma^\rho_{\sigma \lambda} \equiv -\partial_\sigma(\sqrt{-g\,} g^{\sigma \rho}/\sqrt{-g\,})$ and {\it e.g.} $A_{\mu \nu}^{\;\;\;\,;\nu} \equiv \partial^\nu(\sqrt{-g\,} A_{\mu \nu}) /\sqrt{-g\,}$. Without gravitation, Eq. (20) is reduced to the formula

%rownanie 21 
\begin{equation}
(\Box - M^2)\partial^\nu A_{\mu \nu} = - \sqrt{\!f\,}\partial^\nu \left(\varphi F_{\mu \nu} + \zeta\, \bar\psi\, \sigma_{\mu \nu}\, \psi\right)
\end{equation}

\ni equivalent to Eq. (6) for $A_{\mu \nu}$ when this equation is acted on by $\partial^\nu $, where $\partial^\nu A_{\mu \nu} = -\partial^\nu \partial^\nu \chi_\mu = \Box \chi_\mu$ under the condition $\partial^\nu \chi_\nu = 0$ ({\it cf.} Appendix in Ref. [3]). Note that the general field equation (20) for $\chi_\mu$ gets the form
$(...)_{\mu \nu}^{\;\;\;\,;\nu} = -\xi \lambda M^2\chi_\mu $, where $(...)_{\mu \nu}^{\;\;\;\,;\nu} \equiv \partial^\nu[\sqrt{-g\,} (...)_{\mu \nu}] /\sqrt{-g\,}$. If the cosmological constant $\lambda$ is neglected, it is reduced to the field equation $(...)_{\mu \nu}^{\;\;\;\,;\nu}  = 0$ for $A_{\mu \nu}$ or even to $(...)_{\mu \nu} = 0$, when $\cal{L}^{(\chi)}$ is not dependent on $\chi_\mu$ {\it i.e.}, $\cal{L}^{(\chi)} = \cal{L}^{(A)}$ (as in the case of Eq. (14) with $\lambda = 0$). 

Now, making use of the Lagrangian density (14) or field equation (20), we may quantize all fields of "matter"\, in our model, keeping the Einsteinian metric $g_{\mu \nu}$ of spacetime still classical (then, $R_{\mu \nu}$ and $T_{\mu \nu}$ in the Einstein equation (12) or (13) are also classical). In this way, the gravity in our model is considered as classical in its nature.

We can stress such a c-number nature of $g_{\mu \nu}(x)$ by showing that -- after the above quantization of "matter"\,-- the vacuum expectation value of dimensionless quantum pair $\chi_\mu(x)\chi_\nu(x)$ is proportional to the (also dimensionless) metric $g_{\mu \nu}(x)$ of spacetime.

In fact, making use of the repers $e^{\;\rho_0}_\mu(x)\;(\mu = 0,1,2,3$  and $\rho_0 = 0_0,1_0,2_0,3_0)$ orthonormalized as 

%rownanie 22
\begin{equation}
e^{\;\rho_0}_\mu(x) e^{\mu\,\sigma_0}(x) = \eta^{\rho_0\,\sigma_0}\;\;{\rm with}\;\;\eta^{\rho_0\,\sigma_0} = {\rm diag}\;(1,-1,-1,-1) = \eta_ {\rho_0\,\sigma_0} \;,   
\end{equation}

\ni and having the metric $g_{\mu \nu}(x)$ of spacetime defined through the product

%rownanie 23
\begin{equation}
g_{\mu\nu}(x) = e^{\;\rho_0}_\mu(x) e_{\nu\,\rho_0}(x) =  e_\mu^{\;\rho_0}(x) e_\nu^{\;\sigma_0}(x)\eta_{\rho_0\,\sigma_0}\,,
\end{equation}

\ni we can represent the field $\chi_\mu(x)$ in the framework of the metric in the following form:

%rownanie 24
\begin{equation}
\chi_\mu(x) = e_\mu^{\;\rho_0}(x) \chi_{\rho_0}(x) = e_{\mu\,\rho_0}(x) \chi^{\rho_0}(x) \,, 
\end{equation}

\ni  where the projections $\chi_{\rho_0}(x) = e^\mu_{\;\rho_0}(x) \chi_{\mu}(x)$ of the vector field $\left(\chi_{\mu}(x)\right)$ onto the repers $\left(e^{\mu}_{\;\rho_0}(x)\right)$ satisfy the orthogonality condition  

%rownanie 25
\begin{equation}
<\!\!\chi_{\rho_0}(x) \chi_{\sigma_0}(x)\!\!>_{\rm vac} = Z \eta_{\rho_0\,\sigma_0}\,. 
\end{equation}

\ni A positive $Z$ denotes in Eq. (25) a dimensionless normalization factor requiring a cut-off procedure to treat the distribution $<\!\!\chi_{\rho_0}(x) \chi_{\sigma_0}(x')\!\!>_{\rm vac}$ at $x-x' \rightarrow 0$. Here, $e_{\mu\,\rho_0}(x) = \eta_{\rho_0\,\sigma_0} e_\mu^{\;\sigma_0}(x)$\,, $\chi^{\rho_0}(x) = \eta^{\rho_0\,\sigma_0} \chi_{\sigma_0}(x)$, while $\,e_{\mu\,\rho_0}(x) = g_{\mu \nu}(x)e^\nu_{\;\rho_0}(x)$. Then, due to the c-number character of the metric, the following relationship holds between the vacuum expectation value of the pair $\chi_ \mu(x) \chi_ \nu(x)$ and the metric $g_{\mu \nu}(x)$ of spacetime:

%rownanie 26 
\begin{equation}
<\!\!\chi_\mu(x) \chi_\nu (x)\!\!>_{\rm vac} = e_{\mu}^{\;\rho_0}(x) e_{\nu}^{\;\sigma_0}(x) <\!\!\chi_{\rho_0}(x) \chi_{\sigma_0}(x)\!\!>_{\rm vac} = Z g_{\mu \nu}(x)\,.
\end{equation}

The dimensionless value of $Z$, related to the unique status of $\chi_\mu(x) $ in our model,  suggests a distinguished character of the formula (26). For other vector-field structures, {\it e.g.} the electromagnetic vector field $A_\mu(x)$ (of dimension one), we get similar relationships as Eq. (26), but with normalization factors of other dimensions, {\it e.g.} dimension two. Then, {\it e.g.} $<\!\!A_\mu(x) A_\nu (x)\!\!>_{\rm vac}\!\! {\bf :} <\!\!\chi_\mu(x) \chi_\nu (x)\!\!>_{\rm vac}$ is equal to the ratio of both normalization factors of resulting dimension two.

When deriving formally the relationship (26), we considered the metric $g_{\mu \nu}(x)$ as being classical. If the metric of spacetime has to be quantized, then such a relationship ought to be avoided. Thus, as long as $g_{\mu \nu}(x) = Z^{-1}\!\!<\!\!\chi_\mu(x) \chi_\nu (x)\!\!>_{\rm vac}$\,, the gravity remains classical in its nature. In such a case, the c-number Einsteinian metric $g_{\mu \nu}(x)$ of spacetime has to collaborate within a quantum theory without a direct quantization of gravity. Then, our model gets a mixed structure containing a classical as well as a quantum part, the latter dynamically not complete (as yet). Of course, the Einstein equation (12) or (13) is here an independent postulate pertaining to the metric of spacetime. It can be seen that the relationship $<\!\!\!\chi_\mu(x) \chi_\nu (x)\!\!>_{\!{\rm vac}}\! = Z g_{\mu \nu}(x)$ is a solution to the equation $\left(R_{\mu \nu} - \frac{1}{2} g_{\mu \nu} R^\rho_\rho - \lambda g_{\mu\nu} + 8\pi G_N T_{\mu \nu}\right) \!\!<\!\!\chi^\mu \chi^\nu\!\!>_{\!{\rm vac}} = 0$, if and only if the Einstein scalar equation

%rownanie 27
\begin{equation}
R^\rho_\rho + 4\lambda = 8\pi G_N T^\rho_\rho
\end{equation}

\vspace{0.2cm}

\ni holds (following from the Einstein equation (12) or (13) when this is satisfied). In fact, evidently $\left(R_{\mu \nu} - \frac{1}{2} g_{\mu \nu} R^\rho_\rho  - \lambda g_{\mu\nu} + 8\pi G_N T_{\mu \nu}\right) g^{\mu \nu} = -R^\rho_\rho -4\lambda + 8\pi G_N T^\rho_\rho $. 

In the particular case of $\xi = 0$, Lagrangian densities (11) and (14) are identical and do not contain the field $\chi_\mu$ outside of $A_{\mu \nu}$ (being then trivially gauge invariant with respect to $\chi_\mu$). In this case, $\lambda$ does not appear in ${\cal{L}}^{(\chi)} = {\cal{L}}^{(A)}$, though it may be that still $\lambda \neq 0$ in the Einstein equation.

Note that during the formal derivation of relationship (26) no particular dynamics for the vector field $\chi_\mu(x)$ was needed (only its coexistence with classical gravity was required). This makes the relationship (26) an identity for $\chi_\mu(x)$. So, due to the relationship (26), the  metric $g_{\mu\,\nu}(x)$ can be replaced everywhere by the vacuum expectation value ${Z^{-1}\!\!<\!\!\chi_\mu(x) \chi_\nu (x)\!\!>_{\rm vac}}$. Obviously,  the c-number metric $g_{\mu \nu}(x)$ depends functionally on the c-number counterparts of $\chi_\mu(x)$  and other quantum fields of "matter",  contained in the classical energy-momentum tensor $T_{\mu \nu}(x)$ standing on the rhs of Einstein equation (12) or (13). Evidently, the content of such a tensor ought to be consistent (in the sense of the considered classical limit) with our experimental knowledge about cold dark matter and its hypothetical hidden-sector origin.    

Finally, we can see that in our discussion the bold question is round the corner: could the c-number Einstein equation be closely correlated with the quantum part of our model (when it  involves an adequate dynamics for $\chi_\mu(x)$ and other quantum fields), and so be considered as a c-number conclusion from the latter, consistent with {gen{\nolinebreak}eral} relativity? In particular, this correlation should be consistent with the relationship $ g_{\mu \nu}(x)\, = {Z^{-1}\!\!<\!\!\chi_\mu(x) \chi_\nu (x)\!\!>_{\rm vac}}\,$. In the case of affirmative answer to this question, the quantum part of our model would be dynamically complete. Then, physical roots of the classical-in-nature gravity should be traced within the quantum dynamics of $\chi_\mu(x)$ interacting (very weakly) with all quantum fields. 

%\vfill\eject

\vspace{1.7cm}

{\centerline{\bf References}}

\vspace{0.3cm}

\baselineskip 0.73cm

{\everypar={\hangindent=0.65truecm}
\parindent=0pt\frenchspacing

{\everypar={\hangindent=0.65truecm}
\parindent=0pt\frenchspacing

[1]~W.~Kr\'{o}likowski,{\it Acta Phys. Polon.} {\bf B 39}, 1881 (2008); arXiv: 0803.2977 [{\tt hep--ph}]; {\it Acta Phys. Polon.} {\bf B 40}, 111 (2009); {\it Acta Phys. Polon.} {\bf B 40}, 2767 (2009). 

\vspace{0.2cm}

[2]~{\it Cf. e.g.} J. March-Russell, S.M. West, D. Cumberbath and D.~Hooper, {\it J. High Energy Phys.} {\bf 0807}, 058 (2008). 

\vspace{0.2cm}

[3]~W.~Kr\'{o}likowski,  {\it Acta Phys. Polon.} {\bf B 41}, 1277 (2010). 

\vspace{0.2cm}

[4]~{\it Cf. e.g.} L.D.~Landau and E.M.~Lifshitz, {\it The Classical Theory of Fields}, Pergamon Press, New York, 1975; S. Weinberg, {\it Cosmology}, Oxford University Press, Oxford and New York, 2008; V.~Mukhanov, {\it Physical Foundations of Cosmology}, Cambridge University Press, Cambridge, 2005. 

\vspace{0.2cm}

\vfill\eject 

\end{document}